\documentclass[twocolumn,prd,showpacs,superscriptaddress,preprintnumbers,nofootinbib]{revtex4}
\usepackage{graphicx}
\usepackage{epsfig}
\usepackage{bm}
\usepackage{latexsym,amssymb,amsmath,amsfonts,amssymb,txfonts,pxfonts,wasysym,float}
\usepackage{color}

\DeclareBoldMathCommand\blpar{\left(} 
\DeclareBoldMathCommand\brpar{\right)}

\newcommand{\beq}[1]{\begin{equation}\label{#1}}
\newcommand{\eeq}{\end{equation}}
\newcommand{\bea}[1]{\begin{eqnarray} \label{#1}}
\newcommand{\eea}{\end{eqnarray}}
\newcommand{\ba}{\begin{array}}
\newcommand{\ea}{\end{array}}

\def\be{\begin{equation}}
\def\ee{\end{equation}}
\def\gs{\mathrel{
   \rlap{\raise 0.511ex \hbox{$>$}}{\lower 0.511ex \hbox{$\sim$}}}}
\def\ls{\mathrel{
   \rlap{\raise 0.511ex \hbox{$<$}}{\lower 0.511ex \hbox{$\sim$}}}}

\newcommand{\postscript}[2]{\setlength{\epsfxsize}{#2\hsize}
   \centerline{\epsfbox{#1}}}

\newcommand{\comment}[1]{}

\usepackage[usenames,dvipsnames]{xcolor}
\definecolor{orange}{cmyk}{0,0.5,1,0}
\definecolor{rossoCP3}{cmyk}{0,.88,.77,.40}
\definecolor{graa}{rgb}{0.8,0.8,0.8}
\definecolor{blaa}{rgb}{0.2,0.2,0.6}

\begin{document}

\title{\color{rossoCP3} Exploring the superwind mechanism for generating ultrahigh-energy
    cosmic rays using large-scale modeling of starbursts}

\author{Luis A. Anchordoqui}

\affiliation{Department of Physics and Astronomy,  Lehman College, City University of
  New York, NY 10468, USA
}

\affiliation{Department of Physics,
 Graduate Center, City University
  of New York,  NY 10016, USA
}

\affiliation{Department of Astrophysics,
 American Museum of Natural History, NY
 10024, USA
}

\author{Diego F. Torres}

\affiliation{Institute of Space Sciences (ICE-CSIC),  Campus UAB,
  Carrer de Magrans s/n, 08193 Barcelona, Spain}

\affiliation{Instituci\'o Catalana de Recerca i Estudis Avan\c{c}ats
  (ICREA),  E-08010 Barcelona, Spain}

\affiliation{
Institut d'Estudis Espacials de Catalunya (IEEC),
08034 Barcelona, Spain}

\date{April 2020}

\begin{abstract}
  \noindent The Pierre Auger Collaboration has provided a compelling
  indication for a possible correlation between the arrival directions
  of ultrahigh-energy cosmic rays and nearby starburst galaxies. Herein we show how the latest large-scale modeling of starburst galaxies is 
  compatible with the cosmic rays producing the anisotropy signal being accelerated at the terminal shock of superwinds.
\end{abstract}
\maketitle

\section{Introduction}

Starburst-driven superwinds are complex, multi-phase phenomena 
primarily powered
by the momentum and energy injected by massive stars in the
form of supernovae, stellar winds, and
radiation~\cite{Heckman}. According to the book, these superwinds are
ubiquitous in galaxies where the star-formation rate per unit area
exceeds $10^{-1} M_\odot \ {\rm yr}^{-1} \ {\rm kpc}^{-2}$. The
deposition of mechanical energy by supernovae and stellar winds
results in a bubble filled with hot ($T \alt 10^8~{\rm K}$) gas that
is unbound by the gravitational potential because its temperature is
greater than the local escape temperature. The over-pressured bubble
expands adiabatically, becomes supersonic at the edge of the starburst
region, and eventually blows out of the disk into the halo forming a
strong shock front on the contact surface with the cold gas in the
halo. It was conjectured that charged particles can be accelerated by
bouncing back and forth across this terminal shock up to extremely
high
energies~\cite{Anchordoqui:1999cu,Anchordoqui:2018vji}. However, criticisms on
the choice of model parameters characterizing the speed of the shock
and magnetic field strength were reported
in~\cite{Romero:2018mnb,Romero:2019aex,Gutierrez:2020uvk,Muller:2020vdm}.

Two recent results have further promoted a reanalysis, though.
On the one hand, it was recently proven that the 
photo-disintegration of $^4${He} on the cosmic microwave background (CMB) is less severe than earlier thought~\cite{Soriano:2018lly}.
This implies that the physical survival probability of a 10~EeV helium
nuclei coming from the nearest starbursts is $~ 40\%$ larger. On the
other hand, full-blown simulations have been performed to accurately capture the
hydrodynamic mixing and dynamical interactions between the hot and
cold ($T \sim 10^4~{\rm K}$) phases in the outflow~\cite{Schneider},
as well as to provide a precise determination of the magnetic field in
the halo~\cite{Buckman:2019pcj}. Armed with the results of these Monte
Carlo simulations, we reexamine here the acceleration mechanism in
starburst superwinds and show that cosmic ray acceleration up to the
highest observed energies is indeed feasible.

\section{Observational status}
\label{sec:2}

 \begin{figure*}[tb] 
    \postscript{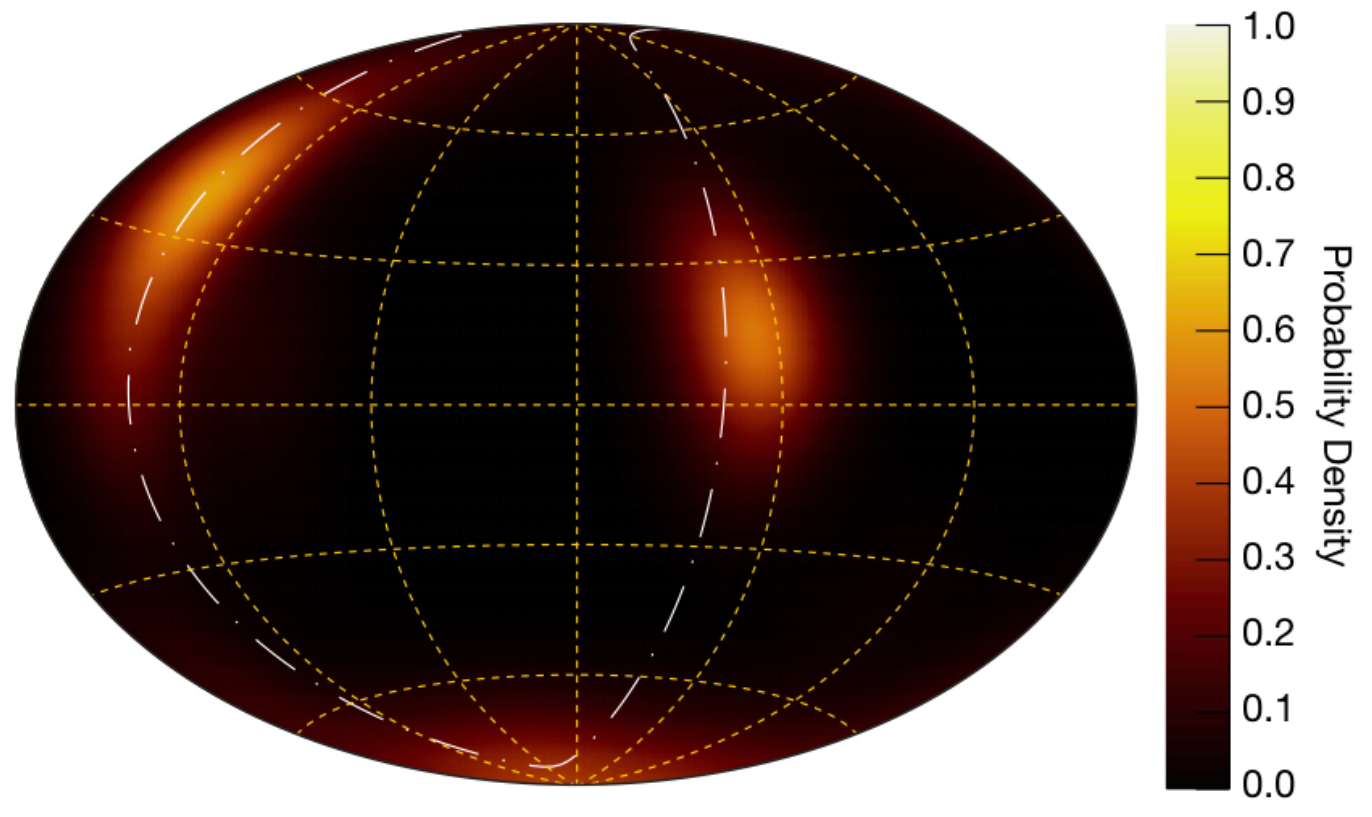}{0.8} 
    \caption{Model flux skymap of UHECR emission from the 23 brightest
      nearby starbursts, with a radio flux larger than 0.3~Jy (selected out
      the 63 objects within 250 Mpc search for $\gamma$-ray emission
      by the Fermi-LAT Collaboration~\cite{Ackermann:2012vca})~\cite{Aab:2018chp}. The continuum emission at
      1.4~GHz has been used as a proxy to weight the UHECR emission and an attenuation factor accounting for energy losses incurred by UHECRs during
      propagation has been included. The map has been smoothed using a Fisher-Von Mises distribution with
      concentration parameter corresponding to a search radius of
      $15.0^\circ$~\cite{Aab:2019ogu}. The color scale indicates
      the probability density of the source skymap as a function of
      position on the sky. The hotspot near the Galactic South Pole
      originates in NGC 253, located at Galactic coordinates ($l =
      97.4^\circ,$ $b = -88^\circ)$; the hotspot concentrated in the
      north-left region of the map is dominated by M82 ($l
      =141.4^\circ,$  $b =
      40.6^\circ$); the hotspot near the Galactic plane on the
      right-side of the map is dominantly produced by NGC 4945
      ($l=305^\circ,$  $b = 13.3^\circ$), 
      but it has a non-negligible contribution from M83
      ($l=314.6^\circ, b = 32^\circ$). The white
      dashed line indicates the Supergalactic Plane. This figure is
      courtesy of Toni Venters. \label{fig:e1}}
\end{figure*}

\begin{figure*}[tb] 
\begin{minipage}[t]{0.49\textwidth}
\postscript{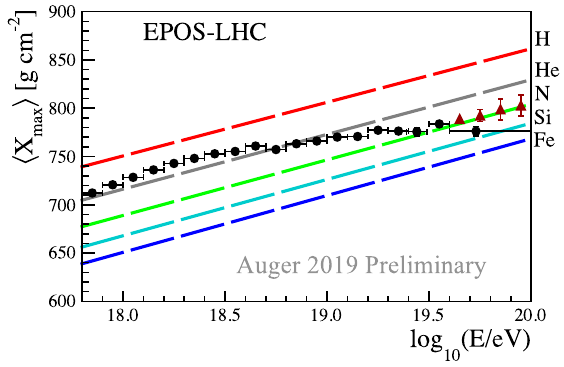}{0.99}
\end{minipage}
\begin{minipage}[t]{0.477\textwidth}
\postscript{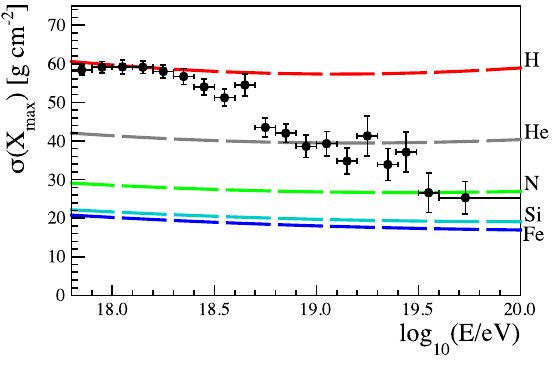}{0.99}
\end{minipage}
\caption{The evolution of $\langle X_{\rm max} \rangle$ (left) and
  $\sigma (X_{\rm max})$ (right) with energy in comparison with the
  predictions of extensive air shower simulations generated using the
  post-LHC hadronic interaction model EPOS-LHC. For details see~\cite{Castellina:2019huz}.
\label{fig:e2}}
\end{figure*}

The Pierre Auger Collaboration reported a $4.5\sigma$-significant
correlation between the arrival direction of cosmic rays with energy
$E > 38~{\rm EeV}$ and a model based on a catalog of bright starburst
galaxies~\cite{Aab:2018chp,Aab:2019ogu}. In the best-fit model,
$11^{+5}_{-4}\%$ of the ultrahigh-energy cosmic-ray (UHECR) flux originates from these objects
and undergoes angular diffusion on a scale
$\theta \sim 15^\circ \pm 5^\circ$. The model signal is largely dominated
by the four nearest starbursts: NGC 253 (which is 2.7~Mpc away), M82
(which is 3.6~Mpc away), and NGC 4945 and M83 (which are both 4~Mpc
away and relatively close to one another); see Fig.~\ref{fig:e1}. M82,
however, is outside the field of view of Auger and therefore the
anisotropy signal is dominated by the other three nearby starbursts. This
is evident in the result of the
test statistics which is almost independent of propagation effects~\cite{Aab:2018chp,Aab:2019ogu}. The angular spread derives from a
Fisher-Von Mises distribution, the equivalent of a Gaussian on the
sphere, and corresponds to a top-hat scale
$\psi \sim 1.59 \times \theta = (24 \pm 8)^\circ$. Median deflections
of particles in the Galactic magnetic field are estimated to be 
\begin{equation}
\theta_{\rm G} \sim 3^\circ \ Z \ \left(\frac{E}{100~{\rm
  EeV}}\right)^{-1} \, ,
  \label{eq:BG}
\end{equation}
where $Z$ is the charge of the UHECR  in
units of the proton charge~\cite{Aartsen:2015dml}. Thus, the requirement $
\theta_{\rm G} \alt \psi$ implies that UHECRs contributing to
the Auger  anisotropy signal should have $Z \alt 10$ and $E/Z \sim
10~{\rm EeV}$.

Measurements of the primary energy, and shower properties like the
atmospheric column depth at which the longitudinal development of a
cosmic-ray shower reaches maximum, $X_{\rm max}$, and its fluctuations
provide a reliable information of the nuclear composition~\cite{Anchordoqui:2018qom}.  The first
two moments of the $X_{\rm max}$ distribution reported by the Pierre
Auger Collaboration, which are displayed in Fig.~\ref{fig:e2}, seem to
indicate that in the energy range of interest the nuclear composition
is consistent with a mixture of CNO, F, and Ne, as required to
accommodate the observed anisotropy signal, i.e., $Z \alt 10$.

First generation UHECR experiments also pointed to a possible
starburst origin for the highest energy
events~\cite{Anchordoqui:2002dj}. With current statistics, the
Telescope Array (TA) Collaboration cannot make a statistically
significant corroboration or refutation of the UHECR
$\leftrightharpoons$ starburst connection~\cite{Abbasi:2018tqo}.
However, TA data have revealed a pronounced directional
hotspot~\cite{Abbasi:2014lda} in arrival directions not far from the
starburst galaxy
M82~\cite{Anchordoqui:2014yva,He:2014mqa,Pfeffer:2015idq}. Indeed, the
latest analysis carried out by the TA Collaboration using 10~yr of
data favors M82 as the origin of the TA
hotspot~\cite{Abbasi:2020fxl}. Constraints based on the isotropic gamma-ray background at $\alt
{\rm TeV}$ measured by the Fermi Large Area Telescope~\cite{TheFermi-LAT:2015ykq} seem to support the
association of UHECRs and nearby starburst galaxies~\cite{Liu:2016brs}. 

Whereas evidence is mounting, a definite answer as to whether a fraction of UHECRs indeed correlates with starburst
galaxies will be given during this decade~\cite{Sarazin:2019fjz}.

\section{Model uptake}

By now, it is well-established that charged particles can gain energy
by diffusing in converging flows~\cite{Fermi}.
Diffusive shock acceleration has been widely accepted as
{\it the} acceleration mechanism for
UHECRs~\cite{Bell:1978zc,Bell:1978fj,Lagage:1983zz,Drury:1983zz,Blandford:1987pw}. 

Consider a steady state collisionless shock propagating
with velocity $u_s >0$ along a plasma at rest with a magnetic field
${\bf B}$ in the direction of the shock normal~\cite{note}. In the shock rest
frame, the flow can be characterized by speeds $u_s$ and $u_s/\zeta$
on different sides of the shock, where $\zeta$ is the density
compression ratio at the shock. For a non-relativistic shock with a
high Mach number, $\zeta=4$. Next, assume that particles of charge
$Ze$ and speed $v \sim c$ start off on the upstream ($u_s$) side of
the shock, and diffuse around through collisions with plasma magnetic
turbulence until they eventually cross the shock and get off
down-stream with lower velocity $u_s/\zeta$.

The diffusion coefficient
$D$ depends on the particle's energy $E$ as well as on the strength 
and structure of the inhomogeneous magnetic field upstream. In the
absence of a fully predictive theory of wave generation and particle
diffusion, it is generally assumed that the diffusion coefficient is
close to the so-called Bohm limit at all energies, defined by
$D = r_L v/3$, where $r_L = E/(ZeB)$ is the Larmor radius of the
particle. This particular value of $D$ is the lowest possible for
isotropic turbulence and can only be achieved if the magnetic field is
disordered on the scale of a particle's Larmor radius (i.e., if
$\lambda/r_L \sim 1$, where $\lambda$ is the mean free path of the
charged particles). Now, in the frame in which the upstream plasma is
at rest the diffusion process isotropizes the angular distribution of
particles. After a period of time $\Delta t = 4 D/(cu_s)$ in which the
particles have diffused in a magneto-hydrodynamic plasma of speed
$u_s$, the particles experience the magnetic turbulence produced by
the downstream plasma. Since the downstream plasma also gravitates to
isotropize the particles elastically, the population of particles
effectively see a plasma moving towards it with speed
$u_s (\zeta - 1)/\zeta$ upon arrival downstream (see
e.g.~\cite{Anchordoqui:2018qom} for details). This process of
quasi-isotropization leads to a net increase in the average particle
speed in the rest frame of the shock interface, yielding a mean
fractional increase in energy of order $\Delta E/E \sim u_s/c$.

Using
the average time that particles spend upstream between shock crossing
we find that the particles accelerate at a rate
$dE/dt \sim \Delta E/\Delta t \sim Eu_s^2/(4D)$. The characteristic
time scale for acceleration to energy $E$ is then
$E/(dE/dt) =4 D/u_s^2$. However, in our back- of-the-envelope estimate
we have overlooked the time spent by the particles downstream. The
downstream diffusion coefficient is expected to be much smaller than
the upstream coefficient, because the downstream magnetic field is
larger due to compression in the shock and probably highly
turbulent. Therefore, inclusion of the downstream dwell time probably
does not increase the acceleration time scale by more than a factor
of 2 and following~\cite{Bell:2013vxa} hereafter we proceed by
assuming that $E/(dE/dt) = 8 D/u_s^2$. In the absence of energy losses
it is straightforward to see that in this approximate model the
maximum energy of the particles is
\begin{equation}
E_{\rm max}  \sim  10 \, Z\left(\frac{u_s}{1000~{\rm km/s}} \right)^2 \left(\frac{\bar
    B}{150\mu{\rm G}} \right) \left(\frac{\tau}{50~{\rm Myr}}\right)~{\rm
    EeV}, 
\label{EmaxDSA}
\end{equation}
  where $\bar B$ is the average magnetic field strength and
  $\tau$ the lifetime of the starburst.

The maximum 
energy of the particles 
is constrained by the Hillas criterium, which states the Larmor radius
of the particle should be no larger than the linear size of the
accelerator, 
\begin{equation}
  E_{\rm max} \alt  Z \left(\frac{\bar
    B}{150\mu{\rm G}} \right) \left(\frac{R_{\rm acc}}{10~{\rm kpc}}\right)~{\rm
  ZeV} \, ,
\label{EmaxH}
\end{equation}
where $R_{\rm acc}$ is the length scale of
acceleration~\cite{Hillas:1985is}. Large-scale optical emission-line and X-ray nebulae oriented
perpendicular to the stellar disks have been observed in several
nearby starburst galaxies, suggesting that the superwind
activity can be traced out to 
$R_{\rm acc} \sim 10~{\rm kpc}$. Indeed, the H$\alpha$ and narrowband
continuum optical images of M82 (taken with the KPNO 0.9~m telescope)
are quite spectacular, with discernible H$\alpha$ line emission extending to
about 11~kpc above the plane of the galaxy; diffuse soft X-ray emission is
seen over the same region by the ROSAT PSPC and
HRI~\cite{Lehnert:1999ra}. The H$\alpha$ line emission has a very
complex morphology that is highly suggestive of outgoing gas, with
loops, tendrils, and filaments of line emission pointing out of the
plane of the galaxy. In addition, XMM–Newton EPIC observations
of M82 show that the soft X-ray emission associated with the superwind
extends out to a height of at least 14~kpc in the North and 7.5~kpc in
the South~\cite{Stevens:2003ek}. In the case of NGC 253, data from
ROSAT PSPC and HRI indicate that the superwind extends out to
10~kpc~\cite{Dahlem}. Near-infrared broad-band and emission-line imaging has revealed the
nucleus of NGC 4945 to be the site of a sizable starburst, the
presence of which is illustrated by the conically shaped starburst
superwind-blown cavity traced at many near-infrared
wavelengths~\cite{Moorwood,Marconi:2000xn}.
However, optical filaments have only been traced
out to about 2~kpc from the plane~\cite{Nakai}. This perhaps could be
justified by the fact that NGC 4945 lies at low Galactic latitude and consequently the
foreground HI column is high. Indeed, as pointed out in~\cite{Strickland:2003xk}, the current lack of evidence for the 10~kpc-scale emission that might
be expected from a starburst-driven superwind, is possibly due to the
resulting high foreground optical depth for H$\alpha$
and soft X-ray radiation. The X-ray data from
Chandra is further biased against a detection of diffuse emission due
to the significantly higher than normal background experienced during
this observation~\cite{Strickland:2003xk}. Finally, M83 has a very complex
morphology as inferred from H$\alpha$ and soft X-ray emission, with an
outflow of hot gas extending roughly $10~{\rm kpc}$
into the halo above the disk of the galaxy~\cite{Ehle:1997cx,Thilker:2004yv}. Altogether,
${\cal O} (10~{\rm kpc})$ scale superwinds have become the de facto
standard while modelling nearby starburst galaxies~\cite{Heckman:1990fe,Weaver:1999nt,
  Strickland:2000dj,Strickland:2009we}.

We now turn to justify the fiducial parameters adopted in
(\ref{EmaxDSA}). The modelling of magnetic field is challenging
  because it is not directly visible to us and we have only a very
  limited idea about the structure of the fields that may exist in the
  acceleration region. Because no measurement technique is capable of
  capturing the entire complexity of the magnetic field, observations
  only reveal hints on $\bar B$. 
  Recently, however, a comprehensive
  study to understand the magnetic field structure in the core and
  halo of the starburst M82 has been carried out~\cite{Buckman:2019pcj}. This study relies on
  the cosmic ray propagation code {\tt GALPROP} that self-consistently model
  the cosmic ray distribution and their diffuse emission.\footnote{\tt
    https://galprop.stanford.edu/} The analysis
  focus on two-dimensional axisymmetric models of the starburst core
  and superwind.

\begin{figure}[tb] 
    \postscript{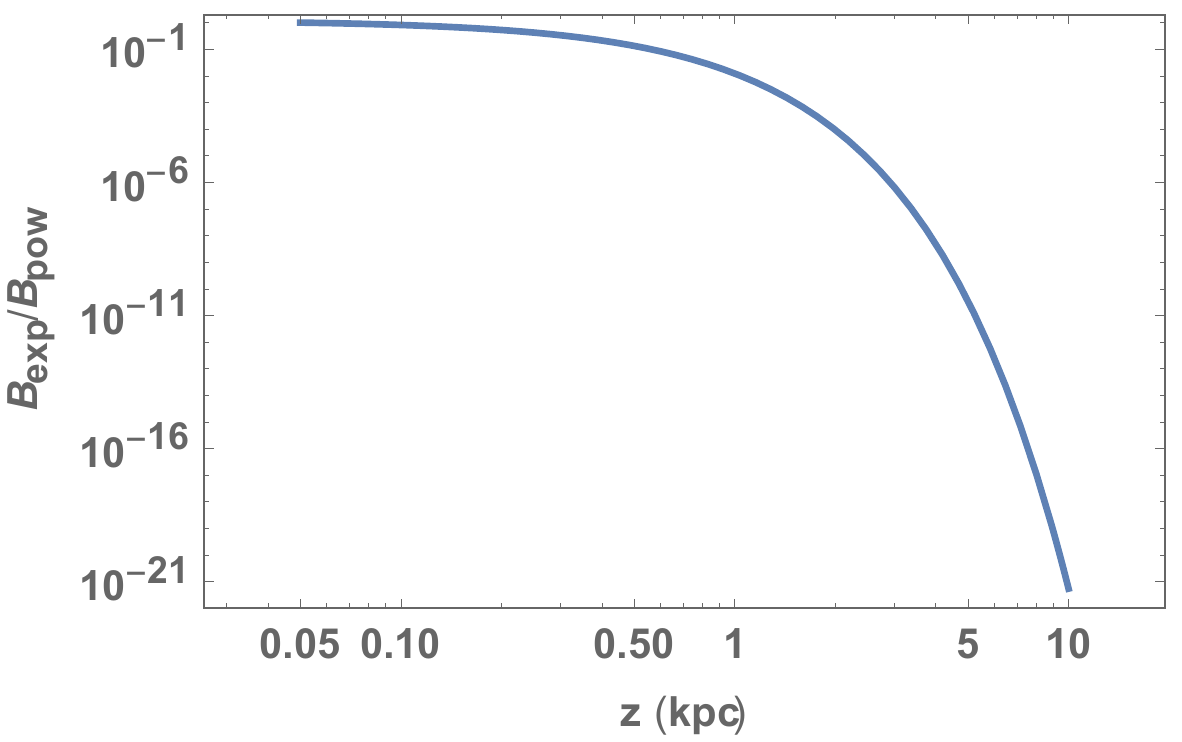}{0.99} 
\caption{The ratio $B_{\rm exp}/B_{\rm pow}$ along the minor axis,
  i.e. for $R=0$ for model ${\cal B}'$.\label{fig:1}}
\end{figure}

\begin{figure}[tb] 
    \postscript{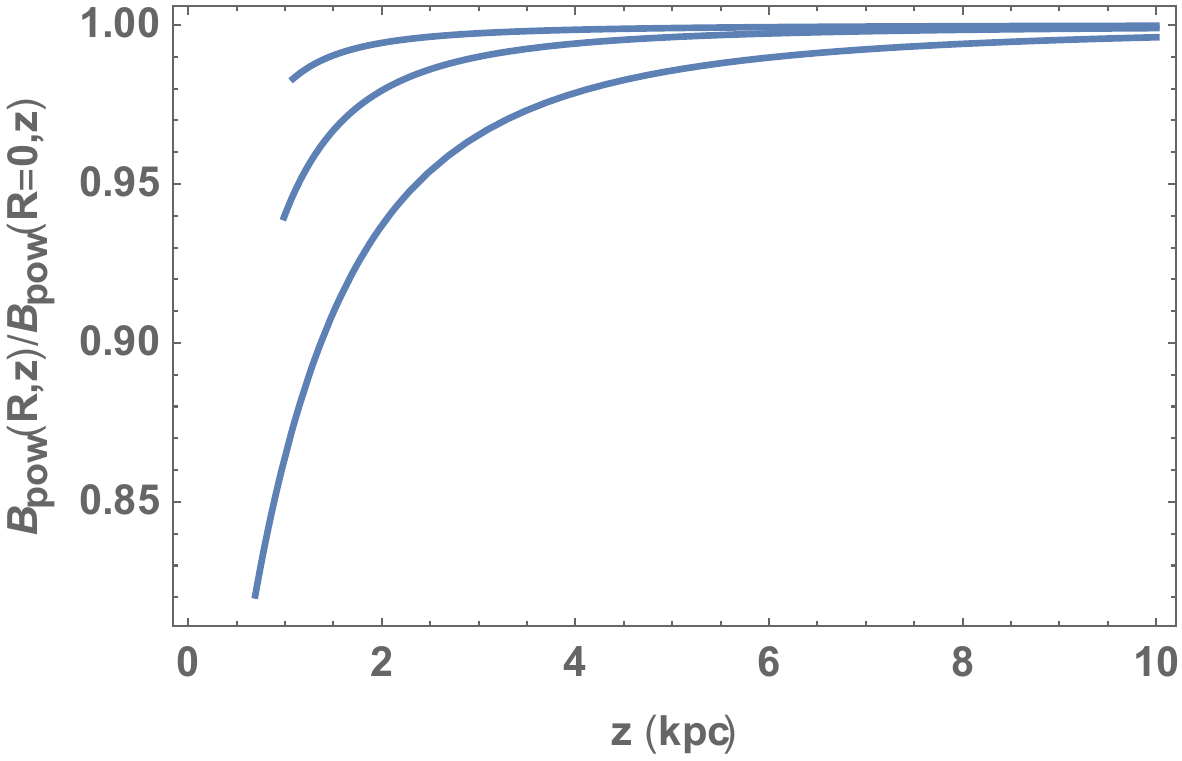}{0.99} 
    \caption{The ratio $B_{\rm pow} (R,z) /B_{\rm pow} (R=0,z)$ for
      model ${\cal B}'$, with $R
      = 0.5, \, 1, \, 2~{\rm kpc}$ downwards. \label{fig:2}}
\end{figure}

The starburst core is modelled as an oblate-ellipsoid, with
cylindrical-radius $R_{\rm core} = 0.2~{\rm kpc}$ and thickness of
$z_{\rm core} = 0.05~{\rm kpc}$.  Throughout the core the
magnetic field strength $B$ and the gas number density $n$ are constants $B_0$ and
$n_0$. Outside the core the field $B({\bf r})$ is defined as ${\rm max} \{B_{\rm
  exp},B_{\rm pow}\}$ where
\begin{equation}
  \frac{B_{\rm exp}}{B_0} = \exp \left(- \frac{r - r_{\rm core}
      (\phi)}{r_{\rm scale} (\phi)} \right) \,,
  \end{equation}
  and
  \begin{equation}
\frac{B_{\rm pow}}{B_0} = \left( \frac{r-z_{\rm core} + z_{\rm
      scale}}{z_{\rm scale}} \right)^{-\beta} \ .
\end{equation}
and where $r = (R^2 + z^2)^{1/2}$ is the spherical radius, with $\phi =
\tan^{-1} |z|/R$, $r_i(\phi) = [(\cos \phi/R_i)^2 + (\sin
\phi/z_i)^2]^{-1/2}$, $z_{\rm scale} = 0.2~{\rm kpc}$, and $i = \{ {\rm core}, {\rm scale} \}$. If
$B_{\rm pow} > B_{\rm exp}$,  along the minor axis (i.e. $R=0$) the magnetic field strength becomes
\begin{equation}
B(z) = B_0 \left(\frac{z-z_{\rm core} + z_{\rm
      scale}}{z_{\rm scale} }\right)^{-\beta} \ .
\end{equation}
To a first approximation we can estimate the average magnetic field as
\begin{eqnarray} \label{average}
  \bar B & \sim & \frac{B_0}{z_{\rm max} - z_{\rm core}} \int_{z_{\rm
      core}}^{z_{\rm max}} \left(\frac{z-z_{\rm core} + z_{\rm
      scale}}{z_{\rm scale} }\right)^{-\beta} \ dz  \\
& = & \frac{B_0 \ z_{\rm scale}^\beta}{(z_{\rm max} - z_{\rm core})
(1-\beta)} \left[(z_{\rm max} - z_{\rm core} + z_{\rm
      scale})^{1-\beta} - z_{\rm scale}^{1-\beta} \right], \nonumber
\end{eqnarray}
with
  $z_{\rm max} = 10~{\rm kpc}$. The normalization constants were determined from a simultaneous fit to
the gamma-ray data from VERITAS~\cite{Acciari:2009wq} and Fermi~\cite{Abdo:2009aa}, as well as the integrated
radio data reported in~\cite{Williams,Adebahr:2012ce,Varenius,Klein}. Using these datasets, the cosmic ray
propagation models were constrained in the parameter plane $(B_0, n_0)$~\cite{Buckman:2019pcj}. There
are three possible combinations (${\cal A}$, ${\cal
    B}$, and ${\cal B}'$) of the two parameters, which provide
excellent fits to the combined data and illustrate unique regions of
the full parameter space. The numerical values for the parameters of models 
${\cal A}$, ${\cal B}$, and ${\cal B}'$ are given in
Table~\ref{tabla}; $u_h$ denotes the maximum assumed superwind
velocity. We note in passing that the values of $B_0$ given in
Table~\ref{tabla} are consistent with
  previous estimates and they are also similar to the magnetic field strength
  observed in
  the nearby starburst galaxy NGC
  253~\cite{Torres:2004,delPozo:2009mh,Paglione:2012ma,DomingoSantamaria:2005qk,Lacki:2013nda}. The  power-law index $\beta$ was inferred from the propagation of the cosmic rays
  out to several kpc, which must create the large radio halo seen at
  22 and 92~cm~\cite{Adebahr:2012ce}. As can be seen in Fig.~\ref{fig:1} the condition $B_{\rm    exp} <B_{\rm pow} $ holds for
  the set of fiducial parameters of model ${\cal B}'$. Substituting for $B_0$ and $\beta$ into (\ref{average}) we obtain $\bar B = 180~\mu{\rm
    G}$.  To visualize the behavior of $B_{\rm pow}$ out of the minor
  axis,  in
  Fig.~\ref{fig:2} we show a comparison of the ratio $B_{\rm pow}
  (R,z) /B_{\rm pow} (R=0,z)$ for various values of $R$. Since $B \geq
  B_{\rm pow}$, we conclude that (\ref{average}) provides a reasonable
  estimate of $\bar B$ for model ${\cal B}'$ in the acceleration
  region.  For models ${\cal A}$ and ${\cal B}$, the values of
  $\bar B$ given in Table~\ref{tabla}
  represent (strictly speaking) lower bounds on the magnetic field strength along the
  minor axis, as $B \geq
  B_{\rm pow}$.  One would have to consider that $B_{\rm exp}$ may exceed $B_{\rm pow}$ for some values of $z$, that is why 
  the estimation using only $B_{\rm pow}$ is a lower limit, although the real value would not exceed this estimation by more than a factor of 2.

  \begin{table}
\caption{Model Parameters ${\cal A}$, ${\cal
    B}$, and ${\cal B}'$. \label{tabla}}
    \begin{tabular}{cccccc}
     \hline
      \hline
      ~~Model~~ & ~~$n_0$ (cm$^{-3}$)~~ & ~~$B_0$ ($\mu$G)~~ &
                                                               ~~$\beta$~~ & ~~$\bar B$ ($\mu$G)~~ & ~~$u_h$ (km/s)~~ \\
      \hline
      ${\cal A}$ &  $\phantom{1}$150  & 150 & 1.0 &    $\phantom{1}$12  & $\phantom{1}$800 \\
      ${\cal B}$  &  $\phantom{1}$675  & 325 & 1.2 & $\phantom{1}$18 &  1000 \\
      ${\cal B}'$ &   1000 & 325 & 0.2 & 180 & 1000 \\
      \hline
      \hline

    \end{tabular}
    \end{table}

Now, the
magnetic field $\bar B$ carries with it an energy density
$\bar B^2/(8\pi)$, and
the flow carries with it an energy flux $> u_s \bar B^2/(8\pi)$. This sets a lower limit on
the rate at which the energy is carried by the out-flowing plasma,
\begin{equation}
  L_B \sim \frac{1}{8} \ u_s \ R_{\rm acc}^2 \ \bar B^2 \,,
\label{eq:LIR}
\end{equation}  
and which must be provided by the source. The flux carried by the
outgoing plasma is a model-dependent parameter, which can be characterized within an
order of magnitude. More concretely, $0.035 \lesssim L_B/L_{\rm IR} \lesssim
0.35$, where $L_{\rm IR} \sim 10^{43.9}~{\rm erg/s}$ is the infrared
luminosity~\cite{Heckman:1990fe}. The lower limit of $L_B$ corresponds to the estimate in~\cite{Heckman:1990fe} considering
a supernova rate of $0.07~{\rm yr}^{-1}$, whereas the upper limit  concurs with 
the estimate in~\cite{Thompson:2006is}, and could be obtained
considering a supernova rate of $0.3~{\rm yr}^{-1}$~\cite{Bregman} while pushing other
model parameters to the most optimistic values. The relation
(\ref{eq:LIR}) yields a magnetic field strength in the range
$15 \lesssim \bar B/\mu{\rm G} \lesssim 150$, which reinforces
the $\bar B$ interval given in Table~\ref{tabla}.
For NGC 4945, the mechanical
luminosity of the superwind is $10^{43.8}~{\rm erg/s}$ and the
supernova rate $> 0.3~{\rm yr}$~\cite{Koornneef}. The IR luminosity of M83 is measured to be $10^{43.2}~{\rm
  erg/s}$~\cite{Vogler:2005bg}. This suggests that both NGC 4945 and
M83 can power a $\mathcal{O} (100~\mu {\rm G})$ magnetic field, too.

Very recently, a series of extremely high resolution global disk simulations of galaxy outflows have been carried out using the Cholla Galactic OutfLow Simulations (CGOLS) program~\cite{Schneider}.
A physically-motivated prescription for clustered supernova served as
a rough proxy for the full-blown simulation of a multiphase outflow
from a disk galaxy. CGOLS is equipped with a two-phase analytic model
capable of fitting the properties of the superwind as a function of
radius. To accurately capture the hydrodynamic mixing and dynamical
interactions between the hot and cold phases in the outflow the
simulations were performed with high resolution ($< 5~{\rm pc}$) across a relatively
large domain (20~kpc).  CGOLS studies favor a volume
average median velocity for the hot phase of the superwind of
$u_{h} \sim 1500~{\rm km/s}$.  CGOLS studies also show that mixing between hot and cold gas in the
wind is an effective way of transferring momentum from one phase to
another and occurs at all radii. This process accelerates cold gas to
$u_{c} \sim 800~{\rm km/s}$ on kpc scales. 

The terminal velocities of the hot ($h$) and cold ($c$) phases of the superwind can be parametrized in terms of
the thermalization efficiency  $\epsilon$  (the fraction of
energy of the central supernovae and stellar winds that goes into the
outflow) and the mass loading factor $\alpha$ (a parameter
commonly used to indicate the relative amount of mass loading; if
$\alpha =1$ the superwind is not mass loaded),
\begin{equation}
\lim_{t \to \infty} u_{h,c} \sim 10^3 \sqrt{\epsilon/\alpha}~~{\rm km/s} \,,
\label{vinfinity}
\end{equation}
where for this order of magnitude calculation, we have assumed that in
total a $100 M_\odot$ star injects ${\cal O} (10^{51}~{\rm erg}$) into
its surroundings during the wind phase~\cite{Heckman}.  Observational
constraints derived from hard X-ray measurements of M82 pin down
medium to high thermalization efficiencies
($0.3 \leq \epsilon \leq 1.0$) and require the volume-filling
superwind hot fluid that flows out of the starburst region to be only
mildly centrally mass loaded
($0.2 \leq \alpha \leq 0.6$)~\cite{Strickland:2009we}. The allowed
phase space then implies a terminal velocity of the M82 superwind in
the range $1400 \leq u_{h} /({\rm km/s}) \leq 2200$, which is consistent with
 CGOLS simulations.  The cold
phase structure of the superwind must be examined using a tracer that
can be directly related to the cold gas.  $\alpha$ was estimated using
observations of $^{12}{\rm CO}$ $\Delta J = 1-0$ transition lines in
NGC 253, obtained with the Atacama Large Millimeter Array
(ALMA)~\cite{Bolatto:2013aqa}. ALMA data favor a mass loading
$\alpha \sim 3$ (with a lower limit of $\alpha \sim 1$). This yields
$600 \alt u_{c} /({\rm km/s})  \alt 1000$, a range which is also 
consistent with CGOLS studies. The estimated superwind velocity of NGC 4945 is somewhat above 1000~km/s~\cite{Chen:1997qn}. Altogether, these numbers justify the
fiducial value, $u_s \sim 1000~{\rm km/s}$, which agrees with the
maximum superwind velocity assumed in the simulations with {\tt
  GALPROP}; see Table~\ref{tabla}~\cite{Buckman:2019pcj}. 

The age of the starbursts is also subject to large
uncertainties. Observations seem to indicate that about 500~Myr ago M82 experienced a tidal
encounter with its large spiral neighbor galaxy, M81, resulting in a
large amount of gas being channelled into the core of the galaxy over
the last 200 Myr, which produced a concentrated
starburst  and an associated pronounced peak in the cluster age
distribution~\cite{deGrijs:2000iv}. This starburst has continued for up to about 50~Myr at
a rate of  $10 M_\odot \ {\rm
  yr}^{-1}$~\cite{deGrijs:2001ec}. A survey of bright evolved stars in
the NGC 253 disk suggest that about 200 Myr ago this galaxy also interacted
with a now defunct companion~\cite{Davidge:2010qg}. Some models 
predict that the age of the starburst in NGC 253  is similar to that
of M82~\cite{Davidge}. Age estimates for the
nuclear starburst of NGC 4945 are sparse and intrinsically uncertain. On the one hand, the
excitation of the starburst core is very low, and this sets a lower
limit on the starburst age of 5~Myr~\cite{Spoon:2000nk}. On the other hand, the
mechanical luminosity associated with its
high supernova rate is consistent with a starburst age of up to
30~Myr or even 50~Myr~\cite{Marconi:2000xn}. The age of the nuclear starburst in M83 was
estimated to be between 10 and 30~Myr by comparing narrowband and
broadband photometry of massive star clusters with theoretical
population synthesis models~\cite{Harris:2001ab}. However, the current
data cannot exclude a longer duration of activity. Altogether this motivated our choice for the starburst age of $\tau \sim 50~{\rm Myr}$.

We now turn to discuss possible energy loss at the source. The accelerating nucleus will gain energy until it eventually suffers
a photodisintegration. The photodisintegration rate depends on the
energy density of the ambient radiation field. This is governed by the
spatial distribution of photons, including both those from the CMB and stellar radiation fields. For compact
regions near the galaxy core, starbursts exhibit an energy density in
their stellar radiation fields which may exceed (or be comparable to)
that of the CMB, but at the superwind scale starlight is expected to
have a negligible energy density compared to that of the
CMB~\cite{Anchordoqui:2007tn,Owen:2019msw}.

\begin{figure}[tb] 
    \postscript{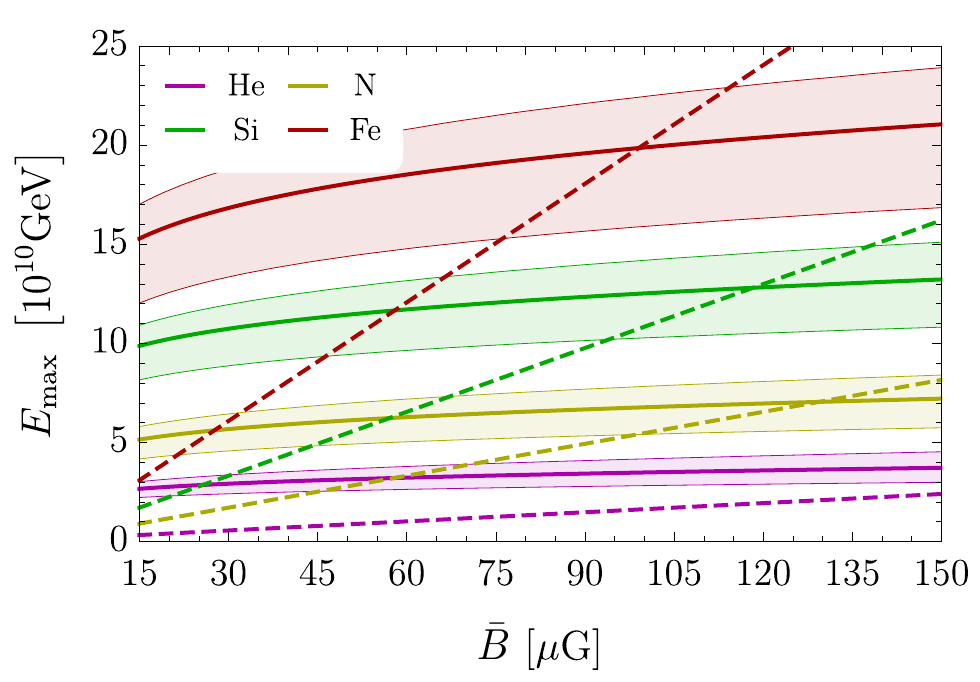}{0.99} 
\caption{Energy loss during the acceleration process. The solid lines
  indicate the time independent cutoff energy for each particle 
  species. This would be a brainwork situation in which the source produces an
  infinite number of identical shocks, with fresh injection at each
  shock and decompression between the shocks, such that the acceleration time-scale $\gg 50~{\rm
    Myr}$~\cite{Anchordoqui:2019mfb,White,Schneider:2,Pope}. It is very unlikely that particles above the cutoff energy
  would have not suffered at least one photodisintegration.  The bands
  encompass 68\% of the nuclei in the statistical sample.  This means
  that only 16\% of the nuclei would reach energies above the upper
  border of the shaded bands and 16\% would only reach energies below
  the lower border. The dashed lines indicate the maximum acceleration
  energy as given by (\ref{EmaxDSA}) in a time period of 50~Myr. This means that
  the maximum energy attained by the particles is the minimum between
  the solid and dashed lines: $E_{\rm max} = {\rm min} \ ({\rm
    solid},\,  {\rm dashed})$ lines, for each particle species. Note
  that for a time scale of 50~Myr, the particles cannot reach extreme
  energies unless the magnetic field is ${\cal O} (100~\mu{\rm
    G})$. For 
  $\tau \sim 50~{\rm Myr}$ and $15 \alt  \bar B/\mu{\rm G} \alt 150$, the photodisintegration energy loss for
  particle species with $Z < 10$ can be neglected, and the maximum
  energy is given by (\ref{EmaxDSA}). \label{fig:5}}
\end{figure}

We duplicate the statistical analysis given in~\cite{Anchordoqui:2019mfb} to determine the
maximum energy at the source. The results for various nuclear species
are shown in Fig.~\ref{fig:5}. We conclude that
for the fiducial value of the acceleration time
scale $\tau \sim 50~{\rm Myr}$ and magnetic field strengths in the
range $15 \alt  \bar B/\mu{\rm G} \alt 150$, the photodisintegration energy
loss  during the acceleration process
can be safely neglected for nuclei with $Z \alt 10$, and therefore (\ref{EmaxDSA})
gives the maximum attainable energy.

Because the sources dominating the anisotropy signal are in our cosmic
backyard, the energy losses during propagation are almost
negligible. This is visible in the result of the test statistics
reported by the Pierre Auger
Collaboration~\cite{Aab:2018chp,Aab:2019ogu}. Note that the mean-free
path of a 70 EeV nitrogen nucleus on the CMB is roughly 3~Mpc~\cite{Anchordoqui:2018qom} and
therefore CNO, F, and Ne can survive the trip from  nearby sources
in the energy range of interest. On the other hand, even if He nuclei
could reach energies beyond 38~EeV in the acceleration process, because of their photodisintegration on the
CMB~\cite{Soriano:2018lly}  they would only contribute to the anisotropy signal near the
minimum energy.

\section{Conclusions}

A number of conclusions are in order connecting back with
(\ref{EmaxDSA}). If we assume the fiducial values for the velocity and
age in (\ref{EmaxDSA}), two of the three models examined (i.e.,
${\cal A}$ and ${\cal B}$, always in the notation
of~\cite{Buckman:2019pcj}) would produce too low averaged values of
energies for the accelerated particles; $E_{\rm max} \sim
Z$~EeV. Unless of course there is observational evidence for $Z$ being
sufficiently high, if models ${\cal A}$ or ${\cal B}$ are realized,
starburst superwinds as a possible origin of the correlation commented
in Sec.~\ref{sec:2} will be challenging. Model ${\cal B}'$, on the
other hand, would naturally lead to particles whose averaged energy
could be in excess of 38 EeV and beyond,
$E_{\rm max} \agt Z \times 10$~EeV.  In this case, relatively light
nuclei (e.g., in the CNO, F, Ne region) would reach very high energies
easily. This is consistent with observation; see
Fig.~\ref{fig:e2}. Even He nuclei may reach the needed UHECR
regime. This would be helped by a less-dramatic photodisintegration
along the way:  new cross section
parameterizations~\cite{Soriano:2018lly} have shown that, e.g., at
$E \sim 10^{1.8}$~EeV and a propagation distance of 3.5 Mpc, the
$^4$He intensity would be 35\% larger than the output of CRPropa 3
program and 42\% larger than the output of SimProp v2r4 program.
Thus, $^4$He could also contribute to the signal near threshold
($E \sim 38~{\rm EeV}$). The latter would of course be easier for
larger velocities: We have assumed a fiducial value of 1000~km/s in
(\ref{EmaxDSA}), but in reality, $u_c < u_s < u_h $ and thus it is
plausible to expect this could be greater than assumed. All in all, we
have shown that UHECR acceleration in superwinds generated by nearby
starbursts (M82, NGC 4945, NGC 253, and M83) is feasible.

Interestingly, we note that if with future data we indeed confirm the correlation 
with starbursts, in the context of superwinds, UHECR observations
would directly feedback on models: The correlation would favor larger
magnetic fields in the outflow. Observational guidance in assessing this conundrum is, at
this point, essential. Tidal
disruption events caused by black holes~\cite{Pfeffer:2015idq}, low-luminosity gamma-ray
bursts~\cite{Zhang:2017moz}, and black hole
unipolar induction~\cite{Gutierrez:2020uvk} have been
identified as potential UHECR accelerators inside starburst
galaxies. However, all of these possible UHECR origins fail to explain which is the inherently
unique feature(s) of starburst galaxies to account for their
correlation~\cite{Anchordoqui:2019ncn}. A true smoking gun for all these proposals would be a correlation with the
distribution of  nearby matter or the more extreme AGNs as opposed to starburst galaxies.
This is not what data seem to suggest. The data yielding the anisotropy signal favor a production
mechanism of UHECRs above 38 EeV which is exclusive to
starbursts. Hence, should future observations confirm the UHECR
$\leftrightharpoons$ starburst connection, Fermi-shock acceleration in
starburst superwinds appears to be a feasible alternative for this distinguishing feature.

\section*{Acknowledgments}
We thank Jorge Fernandez Soriano as well as our colleagues of the POEMMA and Pierre Auger collaborations
for discussion.  We are also grateful to Toni Venters for providing
Fig.~\ref{fig:e1}. The research of L.A.A. is supported by the U.S. National Science Foundation
(NSF Grant PHY-1620661) and the National Aeronautics and Space
Administration (NASA Grant 80NSSC18K0464). The research of D.F.T. is
supported by grants PGC2018-095512-B-I00, AYA2017-92402-EXP, and SGR 2017-1383.

\end{document}